# Thermodynamic Modelling of Phase Equilibrium in Nanoparticles – Nematic Liquid Crystals Composites.

Ezequiel R. Soulé,*[a,b] Linda Reven[c] and Alejandro D. Rey[b]

[a] Institute of Materials Science and Technology (INTEMA), University of Mar del Plata and National Research Council (CONICET), J. B. Justo 4302, 7600 Mar del Plata, Argentina. Email: ersoule@fi.mdp.edu.ar

[b] Department of Chemical Engineering, McGill University, 3610 University st, Quebec H3A2B2, Montreal

[c] Department of Chemistry, McGill University, 801 Sherbrooke St. West, Montreal, Quebec H3A 2K6, Canada

**Abstract.**

*In this work, a theoretical study of phase equilibrium in mixtures of a calamitic nematic liquid crystal and hard spherical nanoparticles is presented. A mean-field thermodynamic model is used, where the interactions are considered to be proportional to the number of contacts, which in turn are proportional to the areas and area fractions of each component. It is shown that, as the radius of the particle is increased, the effect of the particles on the isotropic-nematic transition is less pronounced, and that for large radius the miscibility increases as the particle radius increases.*

## 1. Introduction

Rational combination of nanoparticles (NPs) and a host material opens new ways to the development of advanced materials with applications in optoelectronics, sensing, catalysis, magnetic recording and several other fields [1]. In almost all cases, novel applications strongly depend on the ability to control aggregation and spatial distribution of the particles in the matrix [2-4]. Liquid crystals (LC) have received much attention as dispersing medium for colloidal particles and NPs as a flexible method for generating and controlling self assembly into complex structures [5-9]. LCs are self-organizing anisotropic soft materials that display orientational (nematics) and partial positional order (smectics), widely used in electro-optical applications, and relevant in many biological systems [10-12]. Embeding colloidal particles in a LC matrix can create local distortions in the director orientation field producing an elastic energy that can give rise to short and long range interactions [8,13]. This effect has been shown to result in phase separation and the formation of different structures, like cellular networks [5-7], linear or two-dimensional arrays of particles [13-15] which will strongly affect material properties.

One of the major challenges in nanocomposite research involves creating the means to control the extent of dispersion and self-assembly of the NPs in the host matrix. Understanding the thermodynamic principles that lead to phase separation and phase ordering is then of fundamental importance in the design of new materials, and the ability of predicting phase behaviour and structure formation can help to avoid a trial-and-error procedure that in most cases is used. In the case of micron-sized colloidal particles, the elastic distortions can be strong enough as to suppress Brownian motion, consequently the self-assembly and phase behaviour is dominated by this elastic energy [13,14]. In the case of NPs, as particles are comparable in size with the mesogenic molecules, their capacity to distort the director field is expected to be significantly decreased, and they can "mix" with the LC at a molecular level, so entropic effects must be taken into account to properly describe the phase behaviour of the system. Mixing effects and entropy-driven self assembly (hard-sphere crystallization) are expected to be relevant.

Recently, Matsuyama presented a continuum model for describing phase equilibrium in NP-LC mixtures [16,17]. This model is simple but it captures the basic physics in these systems: it is capable of predicting phase separation in addition to nematic ordering of the LC and colloidal crystallization of the NPs. He analyzed the effects of the different interaction parameters on the transition lines and phase diagrams [16,17]. In this work, we will use an extended version of this model [16,17] to analyze phase behaviour in a mixture of a nematic liquid crystal and nanoparticles. We show that a slight but significant modification of his model is able to capture the fact than when the particles are "macroscopic", in the absence of specific interactions they do not influence the behaviour of the liquid crystal, whereas when the size of the particles are in the nanoscale, they mix at a molecular level with the liquid crystal, producing a dilution effect. Our model predicts that, for large radius, miscibility in the isotropic and in the nematic phase increases when the radius increases, and this result can be used to explain a number of experimental observations.

**2. Model**

The model used in this work is based on a thermodynamic theory proposed by

Matsuyama [16,17], with some significant modifications. We consider the system as composed by a mixture of calamitic nematic LC, and hard spherical NPs. The geometric parameters that characterize the two species are their specific volume, $v$, and their area per unit volume, $a$. For the LC, $v_{LC} = \pi R_{LC}^2 L_{LC}$, $a_{LC} = 2/R_{LC}+2/L_{LC}$, and for the NP, $v_{NP} = 4/3\pi R_{NP}^3$, $a_{NP} = 3/R_{NP}$, where $R_{LC}$ and $L_{LC}$ are the radius and length of a LC molecule and $R_{NP}$ the radius of a nanoparticle. All lengths are given in units of a reference length $l_R$ (the reference volume is defined as $l_R^3$) so they are non-dimensional. The number of LC molecules and NPs in the mixture are $N_{LC}$ and $N_{NP}$, and the total volume is $V$. The dimensionless free energy density $f$ of the mixture is given by four contributions: isotropic mixing free energy ($f_{iso}$), nematic ordering ($f_{nem}$), crystalline ordering ($f_{crys}$) and specific interactions ($f_{int}$):

$$f = \frac{l^3 F}{V R_g T} = f_{iso} + f_{nem} + f_{crys} + f_{int} \tag{1}$$

where $F$ is the total free energy, $R_g$ is the universal gas constant and $T$ the absolute temperature. The isotropic mixing free energy can be approximated by:

$$f_{iso} = \frac{\phi_{LC}}{v_{LC}}\ln(\phi_{LC}) + \frac{\phi_{NP}}{v_{NP}}\ln(\phi_{NP}) + \frac{\phi_{NP}}{v_{NP}}\frac{(4\phi_{NP} - 3\phi_{NP}^2)}{(1-\phi_{NP})^2} + \chi a_p \phi_{NP}\phi_{LC} \tag{2}$$

where $\phi_i = N_i v_i/V$, for $i$= LC, NP, are the volume fractions of each species (the system is considered incompressible so $\phi_{NP} + \phi_{LC} = 1$). The first two terms are the Flory Huggins mixing entropy and the third term is calculated from Carnahan – Starling equation of state for hard spheres and it has been used in the literature to account for the excluded volume of the particles [18-21]. The last term account for isotropic binary interactions and takes into account that the interactions are proportional to the area of contact between LC molecules and NPs, the factor $\phi_{NP}a_{NP}$ is proportional to the total area of NPs and $\varphi_{LC} = \phi_{LC}a_{LC}/(\phi_{LC}a_{LC} + \phi_{NP}a_{NP})$ is the area fraction of liquid crystal and represent the probability that the NP surface is in direct contact with a LC molecule [20]. $\chi = A+B/T$ is the isotropic binary interaction parameter.

The nematic free energy is calculated from the Maier-Saupe theory [22,23]:

$$f_{nem} = \frac{\phi_{LC}}{v_{LC}}\left[-\frac{1}{2}v\varphi_{LC}S^2 - \ln(Z_n) + \frac{3}{2}\Gamma_n S\right] \qquad (3)$$

and is given in terms of the scalar nematic order parameter, $S$, that measures the degree of alignment of LC molecules along a preferential direction. The first term is the orientation-dependent interaction energy, $v$ is the Maier-Saupe quadrupolar interaction parameter and once again the area fraction is used as the probability of contact. $\Gamma_N$ and $Z_N$ are a mean-field parameter and the partition function, given by:

$$S = \frac{1}{Z_N}\int_0^1 \frac{1}{2}(3x^2-1)\exp\left[\frac{\Gamma_N}{2}(3x^2-1)\right]dx \qquad (4)$$

$$Z_N = \int_0^1 \exp\left[\frac{\Gamma_N}{2}(3x^2-1)\right]dx \qquad (5)$$

The logarithm of the partition function can be approximated by a polynomial expression, which is much more efficient from the computational point of view [24]. We approximate it as a sixth-order polynomial in $\Gamma_n$, obtaining the coefficients by a least-squares fitting.

The crystalline free energy is written according to the mean-field model presented by Matsuyama [16,17], and it is analogous to the nematic free energy;

$$f_{cris} = \frac{\phi_{NP}}{v_{NP}}\left[-\frac{1}{2}g\phi_P\sigma^2 - \ln Z_c + \Gamma_c\sigma\right] \tag{6}$$

where the first term accounts for excluded-volume interactions and $g$ is an excluded-volume interaction parameter which for hard spheres is 14.95. Unlike the case of energetic interactions that can be considered to be proportional to the number of contacts, excluded-volume interactions are proportional to the probability that the volume is occupied by the spheres and consequently volume fraction is used. $\Gamma_C$ and $Z_C$ are a mean-field parameter and the partition function, which for a face-centred-cubic structure are given by

$$\sigma = \frac{1}{2\pi Z_c}\int_0^{2\pi}\int_0^{2\pi}\int_0^{2\pi}\cos(x)\cos(y)\cos(z)\exp\left[\Gamma_c\cos(x)\cos(y)\cos(z)\right]dxdydz \tag{7}$$

$$Z_c = \frac{1}{2\pi}\int_0^{2\pi}\int_0^{2\pi}\int_0^{2\pi}\exp\left(\Gamma_c\cos(x)\cos(y)\cos(z)\right)dxdydz \tag{8}$$

As in the nematic case, the crystalline partition function was approximated with a polynomial obtained from a least-squares fitting.

Finally, the last contribution to the free energy is due to specific interactions, following Matsuyama we write this term as:

$$f_{int} = wS^2 a_P \phi_P \varphi_{LC} \tag{9}$$

where $w$ is a binary nematic interaction that accounts for anchoring at the NP surface and distortions in the nematic director at a nanoscale in the vicinity of a NP. We consider these interactions to be proportional to the area of contact and we neglect the coupling interactions between nematic and crystalline ordering.

At equilibrium, the free energy must be a minimum with respect to the order parameters $S$ and $\sigma$. In this conditions, $\Gamma_n = v\varphi_{LC}S - 2wv_{LC}a_{LC}\varphi_{NP}S - cv_{LC}a_{LC}\varphi_{NP}\sigma$ and $\Gamma_c = g\phi_{NP}\sigma - cv_{NP}a_{NP}\varphi_{LC}S$ [16,17]. These expressions can be inserted in eq(1) to calculate the minima.

In addition, the condition for phase coexistence in equilibrium is that the chemical potentials of both components are the same in all the coexisting phases.

## 3. Results and Discussion

As already discussed by Matsuyama [16,17], four different phases can be distinguished based in this mean-field model: isotropic (I) characterized by $S=0$ and $\sigma=0$; nematic (N), where $S>0$ and $\sigma=0$, crystal (C), where $S=0$ and $\sigma>0$, and nematic-crystal (NC), where $S>0$ and $\sigma>0$. Two first-order transition lines can be defined in the $T$-$\phi$ plane: a nematic-isotropic transition (NIT) line, that divides the plane in a region where the equilibrium value of $S$ is 0 and a region where it is greater than 0; and a crystal-isotropic transition (CIT) line, dividing the plane in regions where $\sigma=0$ and $\sigma>0$. Matsuyama calculated phase stability and phase coexistence regions for different values of the interaction parameters [16,17].

Firstly we discuss an important difference between the predictions of our model and Matsuyama's model, regarding the NIT. In his model, the nematic quadrupolar interaction term (first term in the right-had side in equation 3) is written as $v\phi_{LC}^2S^2$, which implies an "effective" quadrupolar interaction parameter $v\phi_{LC}$, while in our model this term is $v\phi_{LC}\varphi_{LC}S^2$, with an effective quadrupolar interaction parameter

$\nu\varphi_{LC}$. This apparently small difference has actually a very significant impact on the nature of the model predictions. The form $\nu\phi_{LC}$, without the use of other specific interactions, would predict that for any size of the particle, the quadrupolar interaction, and consequently the NIT temperature, depends only on the concentration. Nevertheless, as the size of the particle increases, it is expected that, at some point, they start to behave essentially as "macroscopic" solid bodies. In the case that $R_{NP} \gg R_{LC}, L_{LC}$ (when the larger particles are not "nano" anymore but approach micron-sized colloids) the liquid crystal will behave essentially as a pure substance, confined to the free volume between the particles. The particles could affect the properties of the nematic phase by anchoring effects (anisotropic contact energy, distortion of the director field, generation of defects, in our model this the last term in the right-hand side in equation 1) but the quadrupolar interaction in the nematic phase should be identical to that of the pure liquid crystal. In the absence of anchoring effects ($w$=0), the NIT should be not affected by the presence of particles. This is the hypothesis used to model mixtures of submicron colloidal particles and liquid crystals [6,21]. The original model by Matsuyama does not include this behaviour, but our model captures it because $\varphi_{LC}$ approaches 1 when $R_{NP} \gg R_{LC}, L_{LC}$. In figure 1 we show the NIT line for different values of $R_{NP}$ for a system without specific interactions ($w$=0). The straight line represents the "ideal" case, where $a_{LC} = a_{NP}$, as $R_{NP}$ is increased, the line becomes a curve and it shifts towards $T_{NI} = 1$, meaning that infinitely large particles does not affect the NIT temperature. It has to be pointed out that our full model is valid only for true *nano*-particles, with a size in the order of magnitude of molecular size. For micron or sub-micron particles (much larger than a molecule), as discussed before, the liquid crystal will behave as if it was pure so there is no contribution to the mixing energy coming from the liquid crystal, and only the entropy of the "particle gas" remains; consequently the first term in equation 2 should be removed [6,21]. For intermediate cases, a more complex model should be used. Nevertheless, this affects only the calculation of phase coexistence, but the mixing entropy doesn't affect the calculation of the NIT line shown in figure 1.

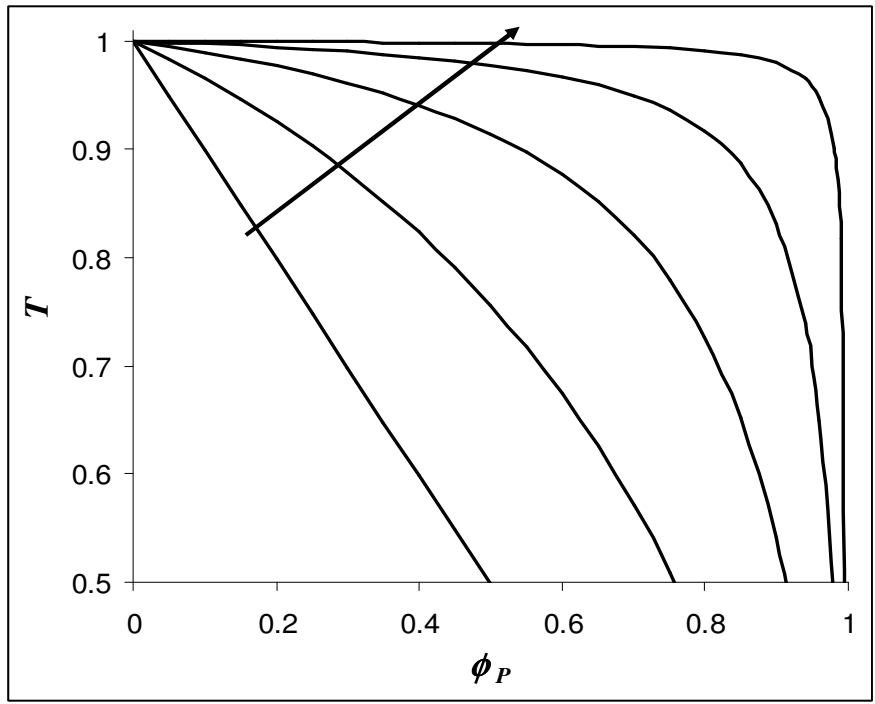

**Fig. 1** NIT lines for a system with $R_{LC}$=0.5, $L_{LC}$=3.5 and $R_{NP}$ = 0.65 (corresponds to $a_{LC} = a_{NP}$), 2, 6, 30 and 300, increasing in the direction of the arrow. Temperatures are normalized with respect to $T_{NI}$ of the pure LC. These lines correspond to a system without specific interactions ($w$ = 0)

Next, we analyze the effect of varying $R_{NP}$ on the phase coexistence for small $R_{NP}$ (in the nanoscale). We will consider the case were the isotropic interaction parameter is high enough for I-I phase coexistence to exist for some values of $R_{NP}$.

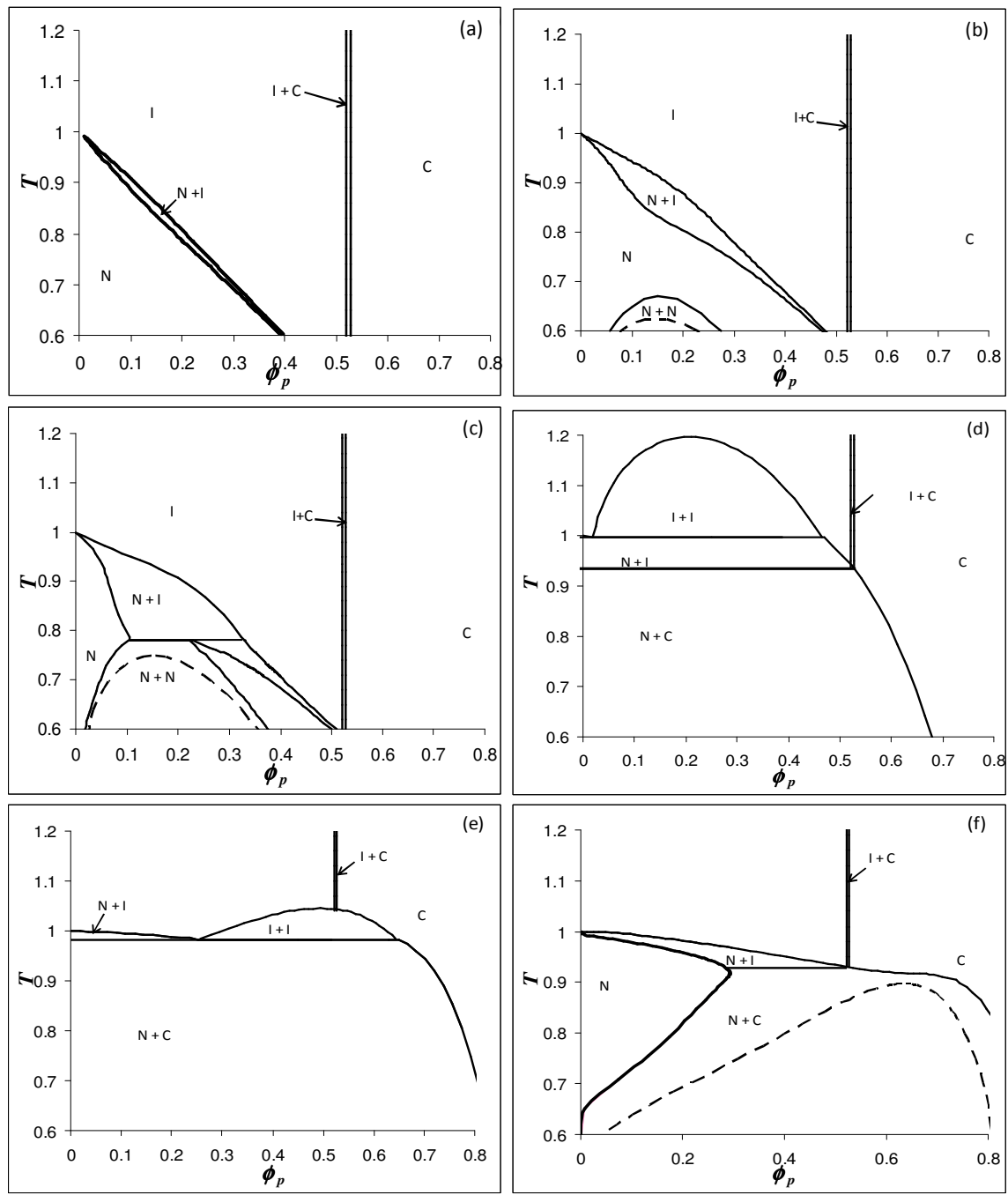

**Figure 2.** Phase diagrams calculated for $R_{LC}$=0.5, $L_{LC}$=3.5 and different values of nanoparticle radius: (a) $R_{NP}$ = 0.65, (b) $R_{NP}$ = 0.9, (c) $R_{NP}$ = 1.2, (d) $R_{NP}$ = 2, (e) $R_{NP}$ = 4, (f) $R_{NP}$ = 6. The interaction parameters are $\chi$=2.5/T and w=0. The dotted line in (b) denotes a metastable L-L phase equilibrium.

Figure 2 shows phase diagrams for different values of $R_{NP}$ (in the same order of magnitude than the LC molecule), for a system with $w = 0$ and $\chi = 2.5/T$. For very small radius (figure 2a) the system is very miscible in the isotropic and nematic states, and only narrow regions of I + N and I + C coexistence exist in the analyzed range of temperature. The NIT line is, as discussed in figure 1, a straight line. For larger radius the I + N regions become broader, and (due to the fact that $\chi > 0$) the metastable I+I equilibrium, buried below the NIT is shifted to higher temperature and can be seen in the range of temperatures analyzed (figure 2b). This buried I+I equilibrium induces a phase separation in the nematic phase, and as a consequence the N + N coexistence region appears. As the NP radius keeps increasing, the metastable I+I coexistence temperature increases so the N + N temperature increases until intercepting the I+N (figure 2c), then it merges completely with the I + N region (not shown in the figures), and finally, a stable I+I coexistence region is observed (figure 2d). The I + I equilibrium temperature is maximum for a given value of $R_{NP}$, further increasing the radius beyond this point starts to shift the I + I region to lower temperatures (figure 2e). This behaviour is produced because miscibility is a balance between entropic effects and enthalpic effects, both of which depend on the radius with a different functionality; as shown in eq. 2, entropy is a function of $1/v_p$ $(1/R_{NP}^3)$ and enthalpy is a function of $a_p$ $(1/R_{NP})$ [20]. For large enough radius, the I + I coexistence becomes metastable again (figure 2f).

An interesting observation is that for large $R_{NP}$, the range of existence of the homogeneous nematic phase becomes broader and the I + N region becomes narrower, so the solubility in the nematic state is increased. This is caused by a combination of the facts that, as discussed before, the mixture is becoming more miscible and the effect of the particles on the NIT becomes less important as the radius increases. This prediction can explain some experimental observations by Qi and Heggman [25]. In their work they analyzed the transition between planar and homeotropic alignment in cells containing LCs doped with NPs. They argued that when the NPs reside at the glass interface, they induce homeotropic alignment at that interface. So, as they outline in their work, the effect of the NPs on the alignment can be explained in terms of solubility in the nematic phase: when the particles are soluble they reside in the bulk and they produce no effect on the surface alignment; when the particles are not very soluble, the excess of particles migrates to the glass interface and induces vertical alignment. In one of their experiments, they analyzed the effect of NPs of different sizes, in a cell with a treated surface that induces planar alignment. For smaller particles, they observed vertical alignment (NP segregation) at NP concentration > 5 % in weight, and planar alignment (complete solubility) at lower NP concentrations, independently of temperature. For larger particles, at a concentration of 10 %, they observed a thermal switch, meaning that the alignment is planar (solubility) at high temperatures and homeotropic (segregation) at low temperature. This implies that the maximum solubility of the smaller NPs in the nematic phase is about 5%, in the whole temperature range analyzed, and large particles in a concentration of 10% are soluble at high temperature, meaning that larger particles are more soluble in the nematic phase than smaller particles. The phase diagrams shown in figure 2 are consistent with behaviour. Obviously a full theoretical description of this phenomenon would require taking into account surface and gradient terms in the free energy, but the trend of the calculated bulk phase diagrams can provide a qualitative explanation for the experimental behaviour. For example, their case of small particles could correspond to a phase diagram similar to figure 2c, where the line separating the N and N + I regions (which represents the maximum solubility in the nematic phase), is essentially vertical (temperature independent); and their case of large particles can be analogous to figure 2f, where the solubility at high temperature is increased, but it is temperature dependent and, when temperature is decreased, the I + N region will be intercepted and particle segregation will be observed.

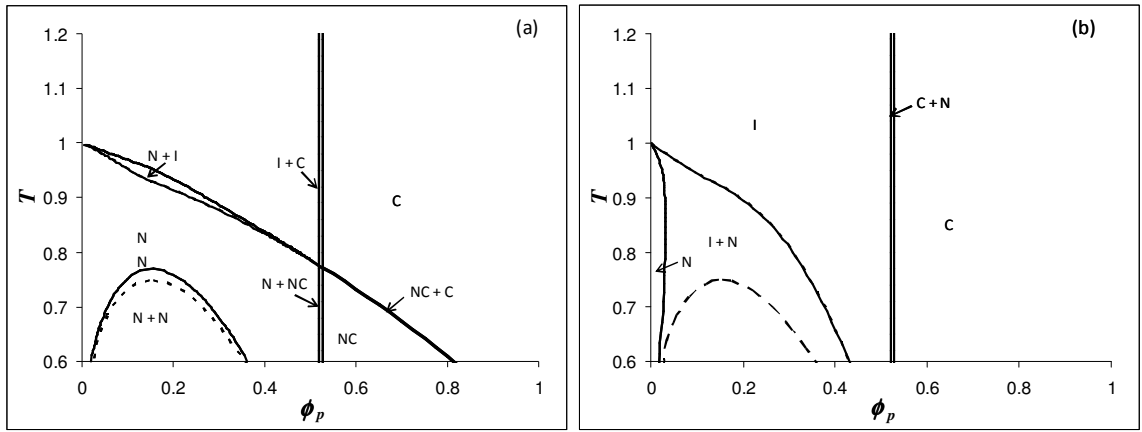

**Figure 3.** Phase diagram for $R_{LC}$=0.5, $L_{LC}$=3.5 $R_{NP}$ = 1.2. The isotropic interaction parameter is $\chi$=2.5/T and the nematic interaction parameter is (a) $w$ = -0.2/T and (b) $w$ = 0.2/T. The dashed line indicates a metastable I + I coexistence.

Figure 3 shows the effect of $w$, the binary nematic interaction parameter, on the phase diagrams (they should be compared to figure 2c that corresponds to the same value of $R_{NP}$, and $w$=0). For a negative value (figure 3a), as shown by Matsuyama [16,17], the NIT line is shifted to higher temperatures and the I + N region becomes narrower, meaning that the particles become more soluble in the nematic phase. The range of existence of the nematic and the nematic-crystal phases is increased, and regions of phase coexistence involving this phase appear in the analysed range. The presence of a buried I + I equilibrium gives rise to a N + N coexistence, as in the case of figures 2b and 2c. For positive values of *w*, (not analyzed by Matsuyama), the opposite happens (figure 3b): the NIT line is shifted to lower temperatures and the I + N coexistence region becomes broader. The range of existence of a homogeneous nematic phase is reduced, as the specific interactions tend to decrease the solubility in the nematic phase. This is as observed in submicron colloidal systems [6,21], where specific interaction arises from elastic distortions on the nematic director produced by the particles, giving rise to a positive free energy that leads to phase separation between a nematic and a isotropic phase. In the case analyzed here, the metastable I + I equilibrium is inside the I + N coexistence region, so no N + N phase coexistence is found. Even if it is completely buried and doesn't affect significantly the shape of the phase diagram, the presence of a metastable I + I equilibrium buried inside the I + N coexistence region can nevertheless have a significant impact on the dynamics of phase separation, leading to the possible formation of double interfaces, as was shown in a previous work [26], so the position of this metastable curve can be relevant for material processing.

## 4. Conclusions

In this work, the phase behaviour of mixtures of spherical nanoparticles and calamitic nematic liquid crystals was analyzed by means of a mean-field thermodynamic model. The model used in this work is an extension of that proposed by Matsuyama [16,17]. We showed that, by considering the nematic quadrupolar interactions to be proportional to the number of contacts between LC molecules (estimated as proportional to the area fraction), the correct effect of the particle radius on the NIT can be reproduced: in the absence of specific interactions (due to anchoring and elastic distortions), as the particles grow in size and become "macroscopic", the effect that they have on the LC becomes less pronounced and the NIT temperature approaches the value corresponding to a pure liquid crystal. The effect of the radius of the particle (in the scale) and the specific interaction (both positive and negative) were analyzed, showing that they affect significantly the phase diagrams. For small particles and negative (attractive) specific interactions, the existence of the homogeneous nematic phase is favoured and a N + N phase coexistence can be induced by a buried metastable I + I equilibrium. As found in previous work for similar systems, the miscibility in the isotropic phase is not a monotonic function of the particle radius: when the particle radius increases, the I + I equilibrium temperature first increases and then decreases. For large enough radius, the NP solubility in the nematic phase is found to increase, a prediction which is consistent with some experimental results found by Qi and Heggman [25].

**References**

[1] Shenhar, R., Norsten, T. B., Rotello, V. M. (2005). *Adv. Mater., 17*, 657.

[2] Vaia, R. A., Maguire, J. F. (2007). *Chem. Mater., 19*, 2736.

[3] Balazs, A. C., Emrick, T., Russell, T. P. (2006). *Science, 314*, 1107.

[4] Krishnamoorti, R. (2007). *MRS Bulletin, 32*, 341.

[5] Cleaver, J., Poon, W. C. K. (2004). *J. Phys. Cond. Matter*, *16*, S1901.

[6] Anderson, V. J., Terentjev, E. M., Meeker, S. P., Crain, J., Poon, W. C. K. (2001). *European Phys. J. E*, *4*, 11.

[7] Petrov, P. G., Terentjev, E. M. (2001). *Langmuir*, *17,* 2942.

[8] Tasinkevych, M., Andrienko, D., (2010) *Cond. Matter Phys.*, *13*, 33603.

[9] Pratibha, R., Park, W., Smalyukh, I. I. (2010) *J. App. Phys.*, *107*, 063511.

[10] Donald, A. M., Windle, A. H., Hanna, S. (2006) *Liquid Crystalline Polymers*, Cambridge University Press, Cambridge, 2nd edn.

[11] Paquet, C., Kumacheva, E. (2008) *Mater. Today*, *11*, 48.

[12] Rey, A. D. (2010). *Soft Matter*, *6,* 3402.

[13] Poulin, P., Weitz, D. A (1998) *Phys. Rev. E*, *58*, 626.

[14] Loudet, J. C., Barois, P., Auroy, P., Keller ,P., Richard, H., Poulin, P. (2004) *Langmuir, 20,* 11336.

[15] Dolganov, P. V., Nguyen, H. T., Joly, G., Dolganov, V. K., Cluzeau, P. (2007) *EPL*, *78*, 66001.

[16] Matsuyama, A., Hirashima, R. (2008). *J. Chem. Phys.*, *128*, 044907.

[17] Matsuyama. A.(2009). *J. Chem. Phys.*, *131*, 204904.

[18] Ginzburg, V. V. (2005) *Macromol., 38,* 2362.

[19] Soulé, E. R., Borrajo, J., Williams, R. J. J. (2007). *Macromol., 40,* 8082.

[20] Soulé, E. R., Hoppe, C. E., Borrajo, J., Williams, R. J. J. (2010) *Ind. Eng. Chem. Res., 49,* 7008.

[21] Popa-Nita, V., van der Schoot, P., Kralj, S. (2006). *European Phys. J. E, 21,* 189.

[22] Maier, W.; Saupe, A. (1959). *Z. Naturforsch., A*, *14*, 882.

[23] Maier, W.; Saupe, A. (1960). *Z. Naturforsch., A*, *15*, 287.

[24] Soulé, E. R., Rey, A. D. (2011). *Liq. Crys.*, *38*, 201.

[25] Qi, H., Heggman, T. (2009). *App. Mater. Int, 8*, 1731.

[26] E. R. Soulé and A. D. Rey 2009 *EPL* **86** 46006